\documentclass{amsart}

\usepackage {epsfig}
\usepackage {graphicx}

 \newtheorem{thm}{Theorem}
 
 \newtheorem{lemma}[thm]{Lemma}
 
 \newtheorem{theorem}[thm]{Theorem}
 \theoremstyle{definition}
 
 \theoremstyle{remark}
 
 \newtheorem{question}[thm]{Question}

 \newcommand{\Real}{\mathbb{R}}

 \newcommand{\Integer}{\mathbb{Z}}

\usepackage{graphicx}
\begin{document}

\title[Rotating waves in Theta neurons]
{Rotating waves in the Theta model for a ring of synaptically
connected neurons}

\author{Guy Katriel}
\address{Einstein Institute of Mathematics, The Hebrew University of Jerusalem, Jerusalem, 91904, Israel}

\email{haggaik@wowmail.com}

\thanks{Partially supported by the Edmund Landau
Center for Research in Mathematical Analysis and Related Areas,
sponsored by the Minerva Foundation (Germany).}


\begin{abstract}
We study rotating waves in the Theta model for a ring of
synaptically-interacting neurons. We prove that when the neurons
are oscillatory, at least one rotating wave always exists. In the
case of excitable neurons, we prove that no travelling waves exist
when the synaptic coupling is weak, and at least two rotating
waves, a `fast' one and a `slow' one, exist when the synaptic
coupling is sufficiently strong. We derive explicit upper and
lower bounds for the `critical' coupling strength as well as for
the wave velocities.  We also study the special case of uniform
coupling, for which complete analytical results on the rotating
waves can be achieved.

\end{abstract}

\maketitle

\section{introduction}
In this work we study rotating waves in rings of neurons described
by the Theta model. The Theta model \cite{ermentrout, ek, hibook,
iz}, which is derived as a canonical model for neurons near a
`saddle-node on a limit cycle' bifurcation, assumes the state of
the neuron is given by an angle $\theta$, with $\theta=(2l+1)\pi$,
$l\in\Integer$ corresponding to the `firing' state, and the
dynamics described by
\begin{equation}\label{thn}
\frac{d\theta}{dt}=1-\cos(\theta)+(1+\cos(\theta))(\beta+I(t)),
\end{equation}
where $I(t)$ represents the inputs to the neuron. When $\beta<0$
this represents an `excitable' neuron, which in the absence of
external input ($I\equiv 0$) approaches a rest state, while if
$\beta>0$ this represents an `oscillatory' neuron which performs
spontaneous oscillations in the absence of external input.

A model of synaptically connected neurons on a continuous spacial
domain $\Omega$ takes the form:
\begin{equation}\label{de}
\frac{\partial \theta (x,t)}{\partial t}=
1-\cos(\theta(x,t))+(1+\cos(\theta(x,t)))\Big[
\beta+g\int_{\Omega}{J(x-y)s(y,t)dy}\Big],
\end{equation}
\begin{equation}\label{syn}
\frac{\partial s (x,t)}{\partial t}+s(x,t)=P(\theta(x,t))(1-c s(x,t)),
\end{equation}
where $J$ is a positive function and $P$ is defined by
\begin{equation}\label{defp}
P(\theta)=\sum_{l=-\infty}^{\infty}{\delta(\theta -(2l+1)\pi)}.
\end{equation}
Here $s(x,t)$ ($x\in\Omega, t\in\Real$) measures the synaptic
transmission from the neuron located at $x$, and according to
(\ref{syn}),(\ref{defp}) it decays exponentially, except when the
neuron fires ({\it{i.e.}} when $\theta(x,t)=(2l+1)\pi$,
$l\in\Integer$), when it experiences a jump. (\ref{de}) says that
the neurons are modelled as Theta-neurons, where the input
$I(x,t)$ to the neuron at $x$, as in (\ref{thn}), is given by
$$I(x,t)=g\int_{\Omega}{J(x-y)s(y,t)dy}.$$
$J(x-y)$ (here assumed to be positive) describes the relative
strength of the synaptic coupling from the neuron at $x$ to the
neuron at $y$, while $g>0$ is a parameter measuring the overall
coupling strength.

The above model, in the case $c>0$, is the one presented in
\cite{ermentrout, oe}. In the case $c=0$ this model is the one
presented in \cite{iz} (Remark 2) and \cite{rubin}. We always
assume $c\geq 0$.

When the geometry is linear, $\Omega=\Real$, it is natural to seek
travelling waves of activity along the line in which each neuron
makes one or more oscillations and then approaches rest. In
\cite{osan} it was proven that for sufficiently strong synaptic
coupling $g$, at least two such waves, a slow and a fast one,
exist, and also that they always involve each neuron firing more
than one time before it approaches rest, while for sufficiently
small $g$ such waves do not exist. It was not determined how many
times each neuron fires before coming to rest, and it may even be
that each neuron fires infinitely many times. Some numerical
results in the case of a one and a two-dimensional geometry were
obtained in \cite{oe}.

In this work we consider a different possibility for the spacial
geometry: $\Omega=S^1$, so the neurons are placed on a ring and
our equations are (\ref{syn}) and
\begin{equation}\label{ge}
\frac{\partial \theta (x,t)}{\partial t}= h(\theta(x,t))+g
w(\theta(x,t))\int_{-\pi}^{\pi}{J(x-y)s(y,t)dy}.
\end{equation}
with
\begin{equation}\label{spec}
h(\theta)=1-\cos(\theta)+\beta(1+\cos(\theta)),\;\;\;w(\theta)=1+\cos(\theta),
\end{equation}
where $J$ is continuous, positive and periodic
\begin{equation}\label{jpos}J(x)>0\;\;\;\forall x\in\Real,
\end{equation}
\begin{equation}\label{jper}
J(x+2\pi)=J(x)\;\;\;\forall x\in\Real,
\end{equation}
and the solutions satisfy the periodicity conditions
\begin{equation}\label{bnt}
\theta(x+2\pi,t)=\theta(x,t)+2\pi m\;\;\;\forall x,t\in\Real
\end{equation}
\begin{equation}\label{bns}
s(x+2\pi,t)=s(x,t)\;\;\;\forall x,t\in\Real
\end{equation}
The integer $m$ (the `winding number') is determined by the
initial condition $\theta(x,0)$, and will be preserved as long as
the solution remains continuous.

In this geometry, a different kind of wave of activity is
possible: a wave that rotates around the ring repeatedly. Such
waves, that is solutions of the form:
\begin{equation}\label{tw1}
\theta(x,t)=\phi(x+vt)
\end{equation}
\begin{equation}\label{tw2}
s(x,t)=r(x+vt)
\end{equation}
where $v$ is the wave velocity, are the focus of our
investigation.

In section \ref{pre} we show that in the case that the winding
number $m=0$, there can exist only trivial rotating waves. Thus
the interesting cases are when $m>0$. Here we study the case
$m=1$, the case $m>1$ being beyond our reach. Thus, this work
concentrates on the first non-trivial case.

Our central results about existence, nonexistence and multiplicity
of rotating waves can be summarized as follows (see figures
\ref{syn1},\ref{syn2} for the {\it{simplest}} diagrams consistent
with these results):

\begin{theorem}
\label{sum}
Consider the equations (\ref{ge}),(\ref{syn}) with conditions
(\ref{bnt}), (\ref{bns}), and $m=1$.

\noindent (I) In the oscillatory case $\beta>0$: for all $g>0$
there exists a rotating wave, with velocity going to $+\infty$ as
$g\rightarrow+\infty$.

\noindent (II) In the excitable case $\beta<0$:

\noindent (i) For $g>0$ sufficiently small there exist no rotating
waves.

\noindent (ii)  for $g$ sufficiently large there exist at least
two rotating waves, a `fast' and a `slow' one, in the sense that
their velocities approach $+\infty$ and $0$, respectively, as
$g\rightarrow +\infty$.
\end{theorem}

Therefore our results bear resemblance to those obtained in
\cite{osan} for the case of a linear geometry. We note that
although for the rotating waves found here each neuron fires
infinitely many times, the reason for this is that it is
re-excited each time, because of the periodic geometry. During
each revolution of the rotating wave, each neuron fires once, so
naively one could think that the analogous phenomenon in a linear
geometry would be a travelling wave with each neuron firing once -
but this was shown to be impossible in \cite{osan}. It is
interesting to note that while in \cite{osan} some restrictions
were made on the coupling function $J$, like being decreasing with
distance, here no such restrictions are imposed beyond
(\ref{jpos}), (\ref{jper}). We would expect however that some
restriction would need to be imposed on $J$ in order to obtain
stability of travelling waves. The whole issue of stability
remains quite open and awaits future investigation. In the case
$\Omega=\Real$, both numerical evidence in \cite{oe,osan} and
results obtained in other models \cite{bres} indicate that the
fast wave is stable while the slow wave is unstable, so we might
conjecture that this is true for the case investigated here as
well - at least under some natural assumptions on $J$. Some
analytical progress on the stability question in the case
$\Omega=\Real$ has recently been achieved in \cite{rubin}.

Let us note that the model considered here, in the case $\beta<0$,
describes waves in an excitable medium, about which an extensive
literature exists (see \cite{winfree} and references therein).
However, most models consider diffusive rather than synaptic
coupling. In the case of the Theta model on a ring, with
{\it{diffusive}} coupling, and $m=1$, it is proven in \cite{er}
that a rotating wave exists regardless of the strength of coupling
({\it{i.e.}} the diffusion coefficient), so that our results
highlight the difference between diffusive and synaptic coupling.

In section \ref{reduction} we reduce the study of rotating waves
to the investigation of the zeroes of a function of one variable.
In section \ref{constant} we investigate the special case in which
the coupling is uniform ($J(x)$ is a constant function), which,
although artificial from a biological point of view, allows us to
obtain closed analytic expressions for the wave-velocity vs.
coupling-strength curves in an elementary fashion. We can thus
gain some intuition for the general case, and obtain information
which is unavailable in the case of general $J$, like precise
multiplicity results. It is interesting to investigate to what
extent the more precise results obtained in the uniform-coupling
case extend to the general case, and we shall indicate several
questions, which remain open, in this direction. In section
\ref{general} we turn to the case of general coupling functions
$J$, and prove the results of theorem \ref{sum} above, obtaining
also some quantitative estimates: lower and upper bounds for the
critical values of synaptic coupling coupling strength $g$, as
well as for the wave velocities.

\section{preliminaries}
\label{pre}

We begin with an elementary calculus lemma which is useful in
several of our arguments below.

\begin{lemma}
\label{cut} Let $f:\Real\rightarrow\Real$ be a differentiable
function, and let $b,c\in \Real$, $b\neq 0$, be constants such
that we have the following property:
\begin{equation}
\label{kp}
f(z)=c \;\;\Rightarrow f'(z)=b.
\end{equation}
Then the equation $f(z)=c$ has at most one solution.
\end{lemma}

\noindent {\sc{proof:}} Assume by way of contradiction that the
equation $f(z)=c$ has at least two solutions $z_0<z_1$. Define
$S\subset \Real$ by
$$S=\{ z>z_0 \;|\; f(z)=c \}.$$
$S$ is nonempty because $z_1\in S$. Let $\underline{z}=\inf S$. By
continuity of $f$ we have $f(\underline{z})=c$. We have either
$\underline{z}>z_0$ or $\underline{z}=z_0$, and we shall show that
both of these possibilities lead to contradictions. If
$\underline{z}>z_0$, then by (\ref{kp}) we have
$$f(z_0)=f(\underline{z})=c$$
$$sign(f'(z_0))=sign(f'(\underline{z}))=sign(b)$$
so we conclude that there exists $z_2\in (z_0,\underline{z})$ with
$f(z_2)=c$, contradicting the definition of $\underline{z}$. If
$\underline{z}=z_0$ then $z_0$ is a limit-point of $S$, which
implies that $f'(z_0)=0$, contradicting (\ref{kp}). These
contradictions conclude our proof.

\vspace{0.4cm}

Turning now to our investigation, we note a few properties of the
functions $h(\theta)$ and $w(\theta)$ defined by (\ref{spec})
which will be used often in our arguments:
\begin{equation}\label{hpi}
h((2l+1)\pi)=2\;\;\;\forall l\in\Integer,
\end{equation}
\begin{equation}\label{hz}
h(2l\pi)=2\beta,\;\;\;\forall l\in\Integer,
\end{equation}
\begin{equation}\label{wz}
w((2l+1)\pi)=0\;\;\;\forall l\in\Integer,
\end{equation}
\begin{equation}\label{wnn}
w(2l\pi)=2\;\;\;\forall l\in\Integer,
\end{equation}

Plugging (\ref{tw1}),(\ref{tw2}) into (\ref{ge}), (\ref{syn}), and
setting $z=x+vt$ we obtain the following equations for $\phi(z)$, $r(z)$:
\begin{equation}\label{teq0}
v\phi'(z)= h(\phi(z))+ g
w(\phi(z))\int_{-\pi}^{\pi}{J(z-y)r(y)dy},
\end{equation}
\begin{equation}\label{req0}
v r'(z)+ r(z)= P(\phi(z))(1-c r(z)).
\end{equation}
In order to satisfy the boundary conditions (\ref{bnt}),(\ref{bns}),
$\phi$ and $r$ have to satisfy
\begin{equation}\label{bnp0}
\phi(z+2\pi)=\phi(z)+2\pi m \;\;\;\forall z\in\Real,
\end{equation}
\begin{equation}\label{bnr0}
r(z+2\pi)=r(z)\;\;\;\forall z\in\Real.
\end{equation}

Let us first dispose of the case of zero-velocity waves, $v=0$.
We get the equations
\begin{equation}\label{teq00}
h(\phi(z))+ g
w(\phi(z))\int_{-\pi}^{\pi}{J(z-y)r(y)dy}=0,
\end{equation}
\begin{equation}\label{req00}
r(z)= P(\phi(z))(1-c r(z)).
\end{equation}
If there exists some $z_0\in\Real$ with $\phi(z_0)=(2l+1)\pi$,
$l\in\Integer$, then, substituting $z=z_0$ into (\ref{teq00}) and
using (\ref{hpi}),(\ref{wz}), we obtain $2=0$, a contradiction.
Hence we must have
\begin{equation}\label{npi}
\phi(z)\neq (2l+1)\pi \;\;\;\forall z\in\Real,\;l\in\Integer
\end{equation}
which implies that $P(\phi(z))\equiv 0$, so that (\ref{req00})
gives $r(z)\equiv 0$, and (\ref{teq00}) reduces to
$h(\phi(z))\equiv 0$, and thus $\phi(z)$ is a constant function,
the constant being a root of $h(\theta)$. This implies, first of
all, that the winding number $m$ is $0$, since a constant
$\phi(z)$ cannot satisfy (\ref{bnp0}) otherwise. In addition the
function $h(\theta)$ must vanish somewhere, which is equivalent to
the condition $\beta\leq 0$. We have thus proven
\begin{lemma}
\label{zvel} Zero-velocity waves exist if and only if $m=0$ and
$\beta\leq 0$, and in this case they are just the stationary
solutions
$$r(z)\equiv 0$$
$$\phi(z)\equiv \pm \cos^{-1}\Big(\frac{\beta+1}{\beta-1}\Big)+2\pi k,\;\;k\in\Integer.$$
\end{lemma}

We will now show that the trivial ``waves" of lemma \ref{zvel} are
the only ones that occur for $m=0$.

\begin{lemma}
\label{mz}
Assume $m=0$.

\noindent (i) If $\beta>0$ there are no rotating waves.

\noindent (ii) If $\beta\leq 0$ the only rotating waves are those
 given by lemma \ref{zvel}.
\end{lemma}

\noindent {\sc{proof:}} Assume $(v,\phi(z),r(z))$ is a solution of
(\ref{teq0}),(\ref{req0}) satisfying (\ref{bnp0}) with $m=0$,
{\it{i.e.}}
\begin{equation}\label{bnp00}
\phi(z+2\pi)=\phi(z) \;\;\;\forall z\in\Real,
\end{equation}
and (\ref{bnr0}). We also assume $v\neq 0$, otherwise we are back
to lemma \ref{zvel}. We shall prove below that $\phi(z)$ must satisfy
(\ref{npi}), and hence that $P(\phi(z))\equiv0$, so that by
(\ref{req0}),(\ref{bnr0}) we have $r(z)\equiv 0$, so that
(\ref{teq0}) reduces to
\begin{equation}
\label{red}
v\phi'(z)=h(\phi(z)).
\end{equation}
Since $v\neq 0$, if $h(\theta)$ has no roots ($\beta>0$),
(\ref{red}) has no solutions satisfying (\ref{bnp00}). If
$h(\theta)$ does have roots ($\beta\leq 0$) then the only
solutions of (\ref{red}) satisfying (\ref{bnp00}) are constant
functions, the constant being a root of $h(\theta)$, and we are
back to the same solutions given in lemma \ref{zvel}, which indeed
can be considered as rotating waves with arbitrary velocity.

It remains then to prove that (\ref{npi}) must hold. Assume by way
of contradiction that $\phi(z_0)=(2l+1)\pi$ for some integer $l$.
By (\ref{hpi}),(\ref{wz}),(\ref{teq0}), and the assumption $v\neq
0$, we have
$$
\phi(z)=(2l+1)\pi \Rightarrow \phi'(z)=\frac{2}{v}.
$$
Thus the assumptions of lemma \ref{cut}, with $f=\phi$,
$c=(2l+1)\pi$, $b=\frac{2}{v}$, are satisfied, and we
conclude that the equation $\phi(z)=(2l+1)\pi$ has at most one
solution, contradicting the fact that, by (\ref{bnp00}), we have
$\phi(z_0)=\phi(z_0+2\pi)=(2l+1)\pi$.

\vspace{0.4cm}

Having found all possible rotating waves in the case $m=0$, we can
now turn to the case $m>0$. In fact, as was mentioned in the
introduction, we shall treat the case $m=1$, the cases $m>1$ being
harder. By lemma \ref{zvel} we know that there are no
zero-velocity waves, so we can assume $v\neq 0$ and define
\begin{equation}\label{dlam}
\lambda=\frac{1}{v},
\end{equation}
so that our equations for the rotating waves can be rewritten
\begin{equation}\label{teq}
\phi'(z)=\lambda h(\phi(z))+\lambda g
w(\phi(z))\int_{-\pi}^{\pi}{J(z-y)r(y)dy},
\end{equation}
\begin{equation}\label{req}
r'(z)+\lambda r(z)=\lambda P(\phi(z))(1-c r(z)),
\end{equation}
with periodic conditions
\begin{equation}\label{bnp}
\phi(z+2\pi)=\phi(z)+2\pi \;\;\;\forall z\in\Real,
\end{equation}
\begin{equation}\label{bnr}
r(z+2\pi)=r(z)\;\;\;\forall z\in\Real.
\end{equation}

\section{Reduction to a one-dimensional equation}
\label{reduction}

We study the equations (\ref{teq}),(\ref{req}) for
$(\lambda,\phi(z),r(z))$ with periodic conditions
(\ref{bnp}),(\ref{bnr}). We will derive a scalar equation (see
(\ref{tt}) below) so that rotating waves are in one-to-one
correspondence with solutions of that equation.

We note first that, since by (\ref{bnr}) we
have $\phi(\Real)=\Real$, and since any rotating wave generates a
family of other rotating waves by translations, we may, without
loss of generality, fix
\begin{equation}\label{fix}
\phi(0)=\pi.
\end{equation}

The following lemma shows that for a rotating wave (in the case
$m=1$) there is at any specific time a unique neuron on the ring
which is firing. This fact is very important for our analysis.

\begin{lemma}
\label{one}
Assume $(\lambda,\phi,r)$ satisfy (\ref{teq}),(\ref{req}) with
conditions (\ref{bnp}),(\ref{bnr}),(\ref{fix}). Then
\begin{equation}
\label{one1}
z\in (0,2\pi) \Rightarrow \pi<\phi(z)<3\pi
\end{equation}
\begin{equation}
\label{one2}
z\in (-2\pi,0) \Rightarrow -\pi<\phi(z)<\pi
\end{equation}
\end{lemma}

\noindent {\sc{proof:}} (\ref{one2}) follows from (\ref{one1}) by
(\ref{bnp}). To prove (\ref{one1}), we first note that certainly
$\lambda\neq 0$, since $\lambda=0$ and (\ref{teq}) imply $\phi(z)$
is constant, contradicting (\ref{bnp}).

We note the key fact that, by (\ref{teq}),(\ref{hpi}) and
(\ref{wz}),
\begin{equation}
\label{kf} \phi(z)=(2l+1)\pi,\;l\in\Integer\;\; \Rightarrow
\;\;\phi'(z)=2\lambda.
\end{equation}
By lemma \ref{cut}, (\ref{kf}) implies that the equation
$\phi(z)=2l+1$ has at most one solution for each $l\in\Integer$.
In particular, since $\phi(0)=\pi$, $\phi(2\pi)=3\pi$ we have
$\phi(z)\neq\pi,3\pi$ for $z\in (0,2\pi)$, and by continuity of
$\phi(z)$ this implies (\ref{one1}).

\vspace{0.4cm} Let us note that if we knew that for rotating waves
the function $\phi(z)$ must be monotone, then lemma \ref{one}
would follow immediately from (\ref{fix}).

\begin{question}
Is it true in general that rotating wave solutions are monotone
(for m>0)?
\end{question}

\begin{lemma}
\label{lampos}
Assume $(\lambda,\phi,r)$ satisfy (\ref{teq}),(\ref{req}) with
conditions (\ref{bnp}),(\ref{bnr}),(\ref{fix}). Then $\lambda>0$.
\end{lemma}

In other words, $v>0$ for all rotating waves with $m=1$, so the
waves rotate clockwise. Of course in the symmetric case $m=-1$ the
waves will rotate counter-clockwise.

\vspace{0.4cm} \noindent {\sc{proof:}} By (\ref{fix}) and
(\ref{kf}) we have $\phi'(0)=2\lambda$. We have already noted that
$\lambda\neq 0$. If $\lambda$ were negative, then $\phi$ would be
decreasing near $z=0$, so for small $z>0$ we would have
$\phi(z)<0$, contradicting (\ref{one1}).

\vspace{0.4cm} Our next step is to solve (\ref{req}),(\ref{bnr})
for $r(z)$, in terms of $\phi(z)$. We will use the following
important consequence of lemma \ref{one}:

\begin{lemma}\label{ret}
$$P(\phi(z))|_{(-2\pi,2\pi)}=\frac{1}{2\lambda}\delta(z)$$
\end{lemma}

\noindent
{\sc{proof:}} By lemma \ref{one} we have
$$P(\phi(z))|_{(-2\pi,2\pi)}=\delta(\phi(z)-\pi)),$$
so we will show that
\begin{equation}
\label{jon}
\delta(\phi(z)-\pi))=\frac{1}{2\lambda}\delta(z).
\end{equation}
Let $\chi(z)\in C_0^{\infty}(\Real)$ be a test function.
Using lemma \ref{one} again we have
\begin{equation}
\label{arg3} \int_{-\pi}^{\pi}{\chi(u)\delta(\phi(u)-\pi)du}=\int_{-\epsilon}^{\epsilon}{\chi(u)\delta(\phi(u)-\pi)du},
\end{equation}
where $\epsilon>0$ is arbitrary. In particular, since
$\phi'(0)=2\lambda>0$, we may choose $\epsilon>0$ sufficiently small
so that $\phi'(z)>0$ for $z\in(-\epsilon,\epsilon)$, so that we
can make a change of variables $\varphi=\phi(u)$, obtaining
\begin{eqnarray}\label{arg4}
\int_{-\epsilon}^{\epsilon}{\chi(u)
\delta(\phi(u)-\pi)du}&=&\int_{\phi(-\epsilon)}^{\phi(\epsilon)}
{\chi(\phi^{-1}(\varphi))\delta(\varphi-\pi)\frac{d\varphi}{\phi'(\phi^{-1}(\varphi))}}
\nonumber\\&=&\frac{\chi(0)}{\phi'(\phi^{-1}(\pi))}=
\frac{\chi(0)}{\phi'(0)}=\frac{\chi(0)}{2\lambda}.
\end{eqnarray}
This proves (\ref{jon}), completing the proof of the lemma.

\vspace{0.4cm}
By lemma \ref{ret} we can rewrite equation (\ref{req}) on the
interval $(-2\pi,2\pi)$ as
\begin{equation}\label{req1}
r'(z)+(\lambda+\frac{c}{2}\delta(z)) r(z)=\frac{1}{2} \delta(z),
\end{equation}
The solution of which is given by
\begin{equation}\label{gs}
r(z)=\Big(\frac{1}{2}H(z)+r(-\pi)e^{-\pi\lambda}\Big)e^{-(\lambda z
+\frac{c}{2}H(z))},
\end{equation}
where $H$ is the Heaviside function: $H(z)=0$ for $z<0$, $H(z)=1$
for $z>0$.
Substituting $z=\pi$ into (\ref{gs}) and using (\ref{bnr}), we
obtain an equation for $r(-\pi)$ whose solution is
$$r(-\pi)=\frac{1}{2}(e^{\pi\lambda+\frac{c}{2}}-e^{-\pi\lambda})^{-1},$$
and substituting this back into (\ref{gs}), we obtain that the
solution of (\ref{req}),(\ref{bnr}) which we denote by
$r_{\lambda}(z)$ in order to emphasize the dependence on the
parameter $\lambda$, is given on the interval $(-2\pi,2\pi)$ by
\begin{equation}\label{fr}
r_{\lambda}(z)=\frac{1}{2}e^{-(\lambda
z+\frac{c}{2}H(z))}\Big[H(z)+(e^{2\pi\lambda+\frac{c}{2}}-1)^{-1}\Big] \;\;\;\;
0<|z|<2\pi.
\end{equation}
We note that, for general $z\in\Real$, $r_{\lambda}(z)$ is given
as the $2\pi$-periodic extension of the function defined by
(\ref{fr}) from $[-\pi,\pi]$ to the whole real line.

The following result, which can be computed from (\ref{fr}), will
be needed later
\begin{equation}
\label{mv}
\int_{-\pi}^{\pi}{r_{\lambda}(u)du}=\frac{1}{2\lambda}\rho_c(\lambda),
\end{equation}
where
$$\rho_c(\lambda)=\frac{e^{2\pi\lambda}-1}{e^{2\pi\lambda+\frac{c}{2}}-1}.$$
We note that
\begin{equation}
\label{rho0}
\rho_0(\lambda)\equiv 1,
\end{equation}
A fact that considerably simplifies the formulas in the case
$c=0$.
\vspace{0.4cm}

The rotating waves correspond to solutions $(\lambda,\phi)$ of the
equation
\begin{equation}\label{twe0}
\phi'(z)=\lambda h(\phi(z))+\lambda g
w(\phi(z))\int_{-\pi}^{\pi}{J(z-y)r_{\lambda}(y)dy},
\end{equation}
with $\phi(z)$ satisfying (\ref{fix}) and
\begin{equation}\label{bphi}
\phi(\pi)=\phi(-\pi)+2\pi.
\end{equation}
To simplify notation, we define
\begin{equation}\label{defrl}
R_{\lambda}(z)=\int_{-\pi}^{\pi}{J(z-y)r_{\lambda}(y)dy},
\end{equation}
so that (\ref{twe0}) is rewritten as
\begin{equation}\label{twe}
\phi'(z)=\lambda h(\phi(z))+\lambda g R_{\lambda}(z) w(\phi(z)).
\end{equation}

We note that (\ref{twe}) is a nonautonomous differential equation
for $\phi(z)$, and since the nonlinearities are bounded and
Lipschitzian, the initial value problem (\ref{twe}),(\ref{fix})
has a unique solution, which we denote by $\phi_{\lambda}$.

Rotating waves thus correspond to solutions $\lambda>0$ of the
equation
\begin{equation}\label{ttt}
\phi_{\lambda}(\pi)-\phi_{\lambda}(-\pi)=2\pi.
\end{equation}
Rewriting (\ref{twe}) and (\ref{ip}) we have
\begin{equation}\label{tww}
\phi_{\lambda}'(z)=\lambda h(\phi_{\lambda}(z))+\lambda g
R_{\lambda}(z) w(\phi_{\lambda}(z)),
\end{equation}
\begin{equation}\label{ip}
\phi_{\lambda}(0)=\pi,
\end{equation}
and defining
\begin{equation}\label{aps}
\Psi(\lambda)=\frac{1}{2\pi}(\phi_{\lambda}(\pi)-\phi_{\lambda}(-\pi)),
\end{equation}
we obtain that rotating waves correspond to solutions $\lambda>0$
of the equation
\begin{equation}\label{tt}
\Psi(\lambda)=1.
\end{equation}

\section{The case of uniform coupling}
\label{constant}

Assuming that the coupling is $J\equiv 1$ we shall be able to
solve for the rotating waves explicitly.
In this case we have, from (\ref{defrl}),(\ref{mv})
$$R_{\lambda}(z)=\int_{-\pi}^{\pi}{r_{\lambda}(y)dy}=
\frac{1}{2\lambda}\rho_c(\lambda),$$ so that (\ref{tww}) reduces
to
\begin{equation}\label{tww0}
\phi_{\lambda}'(z)=\lambda h(\phi_{\lambda}(z))+ \frac{g}{2}
\rho_c(\lambda) w(\phi_{\lambda}(z)).
\end{equation}
The fact that (\ref{tww0}) is an autonomous equation is what makes the treatment of the case
$J$ constant much simpler. Indeed, assume that (\ref{tt}) holds,
so that
\begin{equation}\label{sta}
\phi_{\lambda}(\pi)-\phi_{\lambda}(-\pi)=2\pi.
\end{equation}
Then we have, using (\ref{tww0}), making a change of variables
$\varphi=\phi_{\lambda}(z)$, and using (\ref{sta})
\begin{eqnarray}\label{lam}
1&=&\frac{1}{2\pi}\int_{-\pi}^{\pi}{\frac{\phi'(z)
dz}{\lambda h(\phi(z))+ \frac{g}{2}\rho_c(\lambda)
w(\phi(z))}}\nonumber\\
&=&\frac{1}{2\pi}\int_{\phi_{\lambda}(-\pi)}^{\phi_{\lambda}(\pi)}{\frac{d\varphi}{\lambda h(\varphi)+
\frac{g}{2}\rho_c(\lambda) w(\varphi)}}=
\frac{1}{2\pi}\int_{-\pi}^{\pi}{\frac{d\varphi}{\lambda h(\varphi)+
\frac{g}{2}\rho_c(\lambda) w(\varphi)}}.
\end{eqnarray}
Substituting the explicit expressions for $h$ and $w$ from (\ref{spec}),
and using the formula
\begin{equation}
\label{form}
\frac{1}{2\pi}\int_{-\pi}^{\pi}{\frac{d\phi}{A+Bcos(\phi)}}=\frac{1}{\sqrt{A^2-B^2}}
\;\;\;\;(|A|>|B|),
\end{equation}
(\ref{lam}) becomes
\begin{equation}\label{eqc}
1=\frac{1}{\sqrt{4\lambda^2\beta+2g\lambda\rho_c(\lambda)}},
\end{equation}
so that rotating waves correspond to solutions of (\ref{eqc}),
with their velocities given by $v=\frac{1}{\lambda}$.
We can rewrite (\ref{eqc}) as
\begin{equation}\label{eqc1}
f_{c,\beta}(\lambda)=g,\;\;\;\lambda>0
\end{equation}
where
\begin{equation}\label{df}
f_{c,\beta}(\lambda)=\frac{1-4\beta\lambda^2}{2\lambda\rho_c(\lambda)}.
\end{equation}

In the following lemma we collect some properties of the functions
$f_{c,\beta}(\lambda)$, which are obtained by elementary calculus:

\begin{lemma}\label{prf}
\noindent (i) When $\beta<0$, $f_{c,\beta}$ is positive and convex
on $(0,\infty)$, and
\begin{equation}\label{lz}\lim_{\lambda\rightarrow
0}{f_{c,\beta}(\lambda)}=\infty,
\end{equation}
\begin{equation}\label{li}\lim_{\lambda\rightarrow\infty}{f_{c,\beta}(\lambda)}=\infty.\end{equation}

\noindent (ii) When $\beta\geq 0$, $f_{c,\beta}$ is decreasing on
$(0,\infty)$, and (\ref{lz}) holds. If $\beta>0$ it has a zero at
$\lambda=\frac{1}{2\sqrt{\beta}}$, if $\beta=0$ it is positive on
$(0,\infty)$ and
$\lim_{\lambda\rightarrow\infty}{f_{c,0}(\lambda)}=0$.
\end{lemma}

From lemma \ref{prf} we conclude that when $\beta<0$ (\ref{eqc1}) has exactly
two solutions if $g>\Omega(c,\beta)$, where
\begin{equation}\label{defom}
\Omega(c,\beta)=\min_{\lambda>0}{f_{c,\beta}}(\lambda),
\end{equation}
which we will denote by
$$\underline{\lambda}_{c,\beta}(g)<\overline{\lambda}_{c,\beta}(g),$$
no solution if $g<\Omega(c,\beta)$, and a unique solution when
$g=\Omega(c,\beta)$.

When $\beta\geq 0$, part (ii) of lemma \ref{prf} implies that
(\ref{eqc1}) has a unique solution for any $g>0$, which we denote
by $\lambda_{c,\beta}(g)$.

An elementary asymptotic analysis of the equation
(\ref{eqc1}) yields
\begin{lemma}
\label{asa}
\noindent
(i) When $\beta<0$ we have the following asymptotics as $g\rightarrow \infty$
\begin{equation}\label{asl1}
\overline{\lambda}_{c,\beta}(g)=\frac{e^{-\frac{c}{2}}}{2|\beta|}g+O\Big(\frac{1}{g}\Big)\;\;\;as\;\;g\rightarrow\infty.
\end{equation}
For $\underline{\lambda}_{c,\beta}(g)$, in case $c>0$ we have
\begin{equation}\label{asl2}
\underline{\lambda}_{c,\beta}(g)=\frac{1}{2}\sqrt{\frac{1}{\pi}(e^{\frac{c}{2}}-1)}\frac{1}{\sqrt{g}}+O\Big(\frac{1}{g}\Big)\;\;\;as\;\;g\rightarrow\infty,
\end{equation}
while in case $c=0$ we have
\begin{equation}\label{asl3}
\underline{\lambda}_{0,\beta}(g)=\frac{1}{2}\frac{1}{g}+O\Big(\frac{1}{g^2}\Big)\;\;\;as\;\;g\rightarrow\infty.
\end{equation}

\noindent (ii) For $\beta\geq 0$, the asymptotics of
$\lambda_{c,\beta}(g)$ as $g\rightarrow\infty$ are the same as
those of $\underline{\lambda}_{c,\beta}(g)$, given in (\ref{asl2})
for $c>0$ and (\ref{asl3}) for $c=0$.
\end{lemma}

We thus obtain
\begin{theorem}\label{unij} When $J\equiv 1$:

\noindent (I) In the excitable case $\beta< 0$:

\noindent (i) If $g>\Omega(g,c)$ there exist two rotating waves with
velocities given by
\begin{equation}
\label{vdef}
\underline{v}_{c,\beta}(g)=\frac{1}{\overline{\lambda}_{c,\beta}(g)},\;\;\;\overline{v}_{c,\beta}(g)=\frac{1}{\underline{\lambda}_{c,\beta}(g)},
\end{equation}
and we have, for the slow wave
\begin{equation}\label{asv1}
\underline{v}_{c,\beta}(g)=2|\beta|e^{\frac{c}{2}}\frac{1}{g}+O\Big(\frac{1}{g^3}\Big)\;\;\;as\;\;g\rightarrow\infty,
\end{equation}
for the fast wave when $c>0$:
\begin{equation}\label{asv2}
\overline{v}_{c,\beta}(g)=2\sqrt{\frac{\pi}{e^{\frac{c}{2}}-1}}\sqrt{g}+O(1)\;\;\;as\;\;g\rightarrow\infty.
\end{equation}
while for the fast wave when $c=0$
\begin{equation}\label{asv22}
\overline{v}_{0,\beta}(g)=2g+O\Big(\frac{1}{g}\Big)\;\;\;as\;\;g\rightarrow\infty.
\end{equation}
\noindent (ii) If $g=\Omega(c,\beta)$ there exists a unique rotating wave with
velocity
\begin{equation}\label{asv3}
v=\underline{v}_{c,\beta}(g)=\overline{v}_{c,\beta}(g).
\end{equation}
\noindent (iii) If $g<\Omega(c,\beta)$ there exist no rotating waves.

\noindent (II) When $\beta\geq 0$, there exists a unique rotating
wave for any $g>0$, whose velocity is given by
\begin{equation}
\label{vdef1}v_{c,\beta}(g)=\frac{1}{\lambda_{c,\beta}(g)},
\end{equation}
and for large $g$ it has the same asymptotics as in
(\ref{asv2}),(\ref{asv22}) in the cases $c>0$, $c=0$,
respectively.
\end{theorem}

In the excitable case we thus have two rotating waves born at a
supercritical saddle-node bifurcation as the coupling strength $g$
crosses $\Omega(c,\beta)$.

\vspace{0.4cm} We now note that in the special case $c=0$ (the
model introduced in \cite{iz}) we can obtain more explicit
expressions. Using (\ref{rho0}) we have
$$f_{0,\beta}(\lambda)=\frac{1-4\beta\lambda^2}{2\lambda}.$$
The minimum in (\ref{defom}) can now be computed explicitly, and we obtain, when
$\beta<0$,
$$\Omega(0,\beta)=2\sqrt{|\beta|}.$$
We can also solve (\ref{eqc1}) explicitly, and obtain the
velocities of the rotating waves. When $\beta<0$,
$g>\Omega(0,\beta)$
$$\underline{v}_{0,\beta}(g)=g-\sqrt{g^2+4\beta},\;\;\;\overline{v}_{0,\beta}(g)=g+\sqrt{g^2+4\beta}.$$
When $\beta\geq 0$, for all $g>0$
$$v_{0,\beta}(g)=\sqrt{g^2+4\beta}+g.$$

\begin{figure}
\centering
    \includegraphics[height=7cm,width=7cm, angle=270]{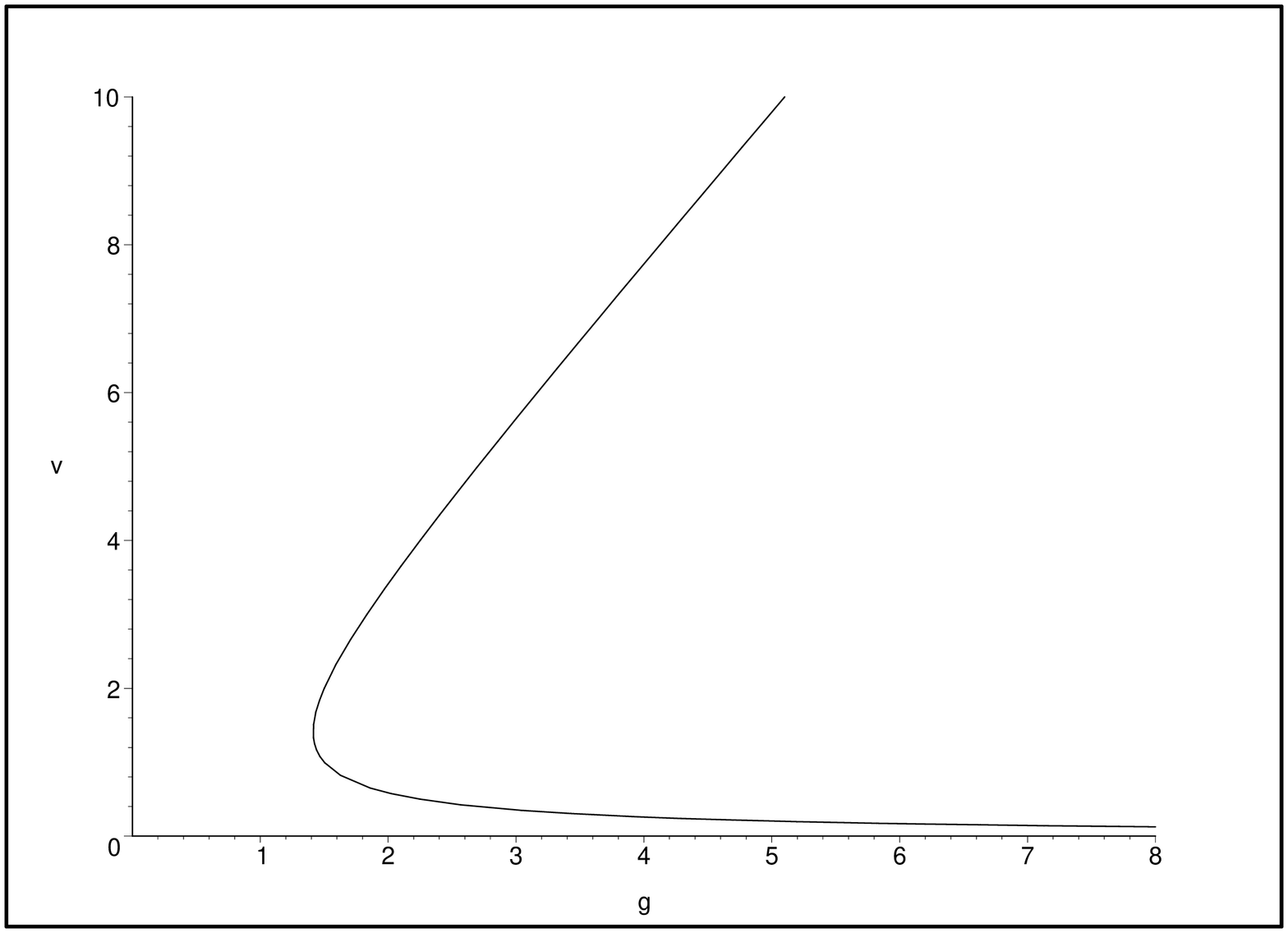}
    \caption{Velocity of waves ($v$) vs. coupling strength ($g$) for the case $J\equiv 1$, $c=0$, $\beta=-0.5$.}
    \label{syn1}
\end{figure}

\begin{figure}
\centering
    \includegraphics[height=7cm,width=7cm, angle=270]{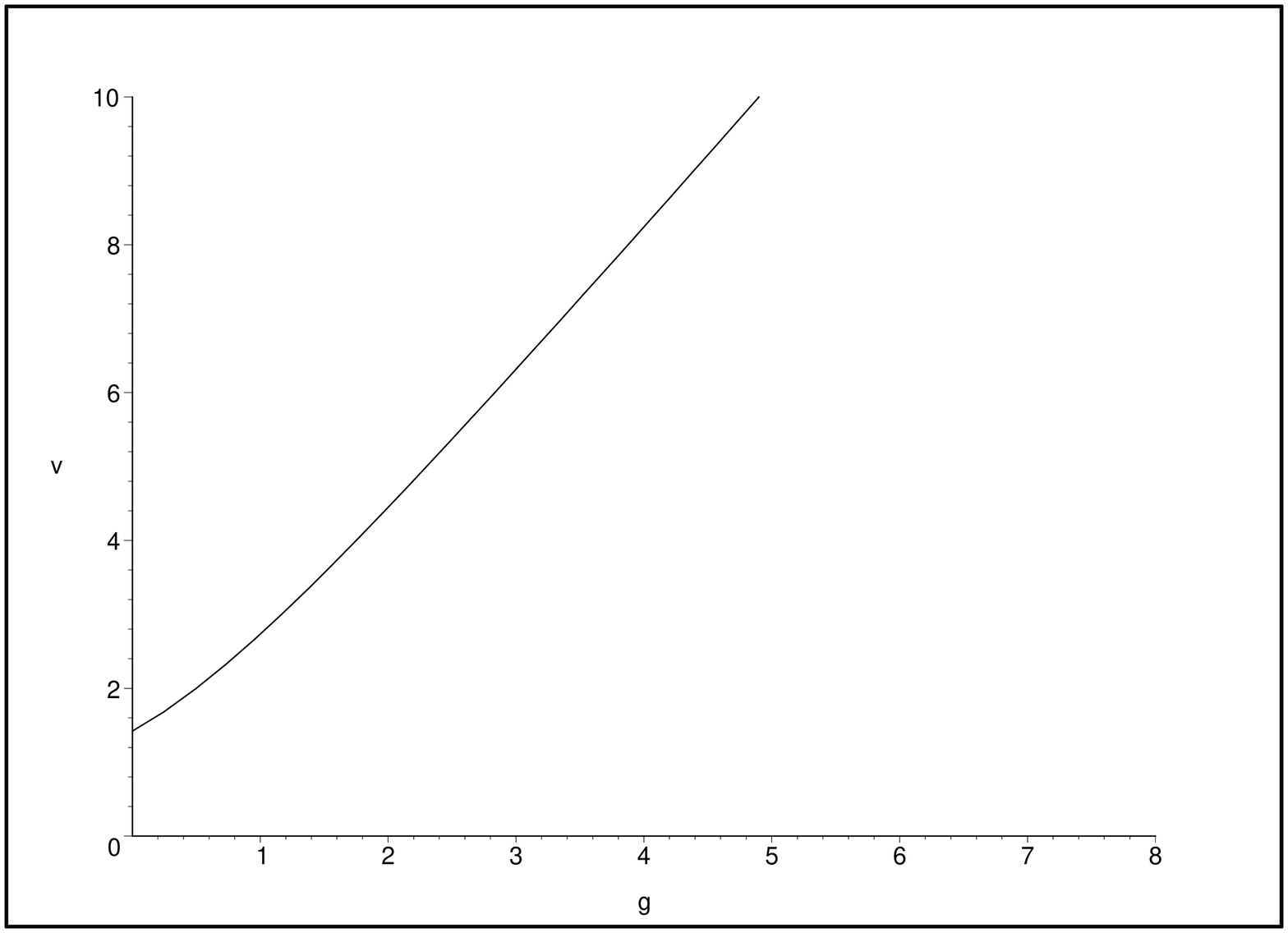}
    \caption{Velocity of waves ($v$) vs. coupling strength ($g$) for the case $J\equiv 1$, $c=0$, $\beta=0.5$.}
    \label{syn2}
\end{figure}

\begin{figure}
\centering
    \includegraphics[height=7cm,width=7cm, angle=270]{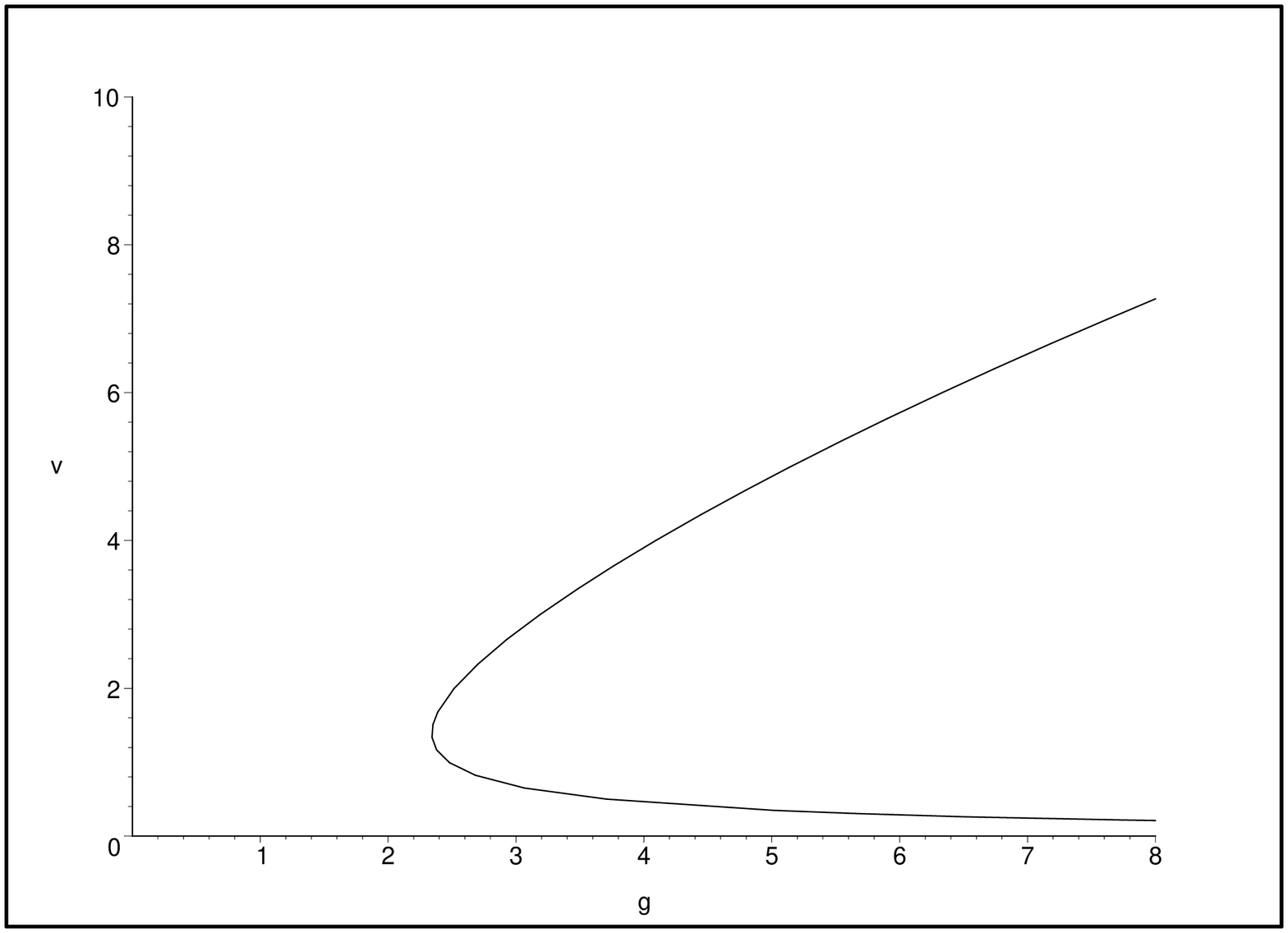}
    \caption{Velocity of waves ($v$) vs. coupling strength ($g$) for the case $J\equiv 1$, $c=1$, $\beta=-0.5$.}
    \label{syn3}
\end{figure}

\begin{figure}
\centering
    \includegraphics[height=7cm,width=7cm, angle=270]{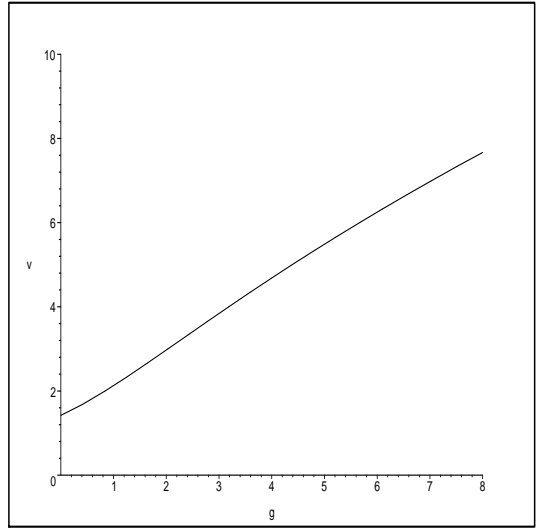}
    \caption{Velocity of waves ($v$) vs. coupling strength ($g$) for the case $J\equiv 1$, $c=1$, $\beta=0.5$.}
    \label{syn4}
\end{figure}

\vspace{0.4cm} Figures \ref{syn1},\ref{syn2} show the
wave-velocity vs. coupling strength diagrams for the rotating
waves when $J\equiv1$, $c=0$, in an excitable ($\beta=-0.5$) and
an oscillatory ($\beta=0.5$) case. In figures
\ref{syn3},\ref{syn4} we change $c$ to $c=1$.

\section{The general case}
\label{general}

We now return to the case when $J$ is a general continuous
positive $2\pi$-periodic function, and prove that several of the
results about rotating waves obtained above for the special case
$J\equiv 1$ remain valid, though the proofs are necessarily less
direct.

\begin{lemma}\label{psiprop}
$$\lim_{\lambda\rightarrow 0}{\Psi(\lambda)}=0.$$
\end{lemma}

\noindent {\sc{proof:}} We shall prove that
\begin{equation}\label{phi0}
\phi_{\lambda}(z)=\pi+O(\lambda) \;\;\;as\;\; \lambda\rightarrow 0
\end{equation}
uniformly in $z\in [-\pi,\pi]$. The lemma follows immediately from
this and from (\ref{aps}).

When $c>0$, the claim (\ref{phi0}) is immediate, since, using
(\ref{fr}),
$$\lim_{\lambda\rightarrow 0}{r_{\lambda}(z)}=\frac{1}{2}e^{-\frac{c}{2}H(z)}
\Big[(e^{\frac{c}{2}}-1)^{-1}+H(z)\Big],\;\;\;0<|z|<2\pi,$$
so that
\begin{equation}
\label{rbig0}
\lambda R_{\lambda}(z)=O(\lambda)\;\;\;as\;\;\lambda\rightarrow 0,
\end{equation}
which implies that the right-hand side of (\ref{twe}) is $O(\lambda)$.

For $c=0$, $r_{\lambda}(z)$ becomes
singular as $\lambda\rightarrow 0$, so we need a
more refined argument. For $\lambda r_{\lambda}(z)$
we have
\begin{equation}\label{propr}
\lambda r_{\lambda}(z)=\frac{1}{4\pi}
+O(\lambda)\;\;\;as\;\;\lambda\rightarrow 0,
\end{equation}
uniformly in $z\in [-\pi,\pi]$, hence
\begin{equation}
\label{rbig}
\lambda R_{\lambda}(z)=\frac{1}{2}\overline{J}
+O(\lambda)\;\;\;as\;\;\lambda\rightarrow 0,
\end{equation}
where
$$\overline{J}=\frac{1}{2\pi}\int_{-\pi}^{\pi}{J(x)dx}.$$
The standard theorems on
dependence of solutions of initial-value problems on parameters
hence imply
$$\phi_{\lambda}(z)=\phi_0(z)+O(\lambda)\;\;\;as
\;\;\lambda\rightarrow 0,$$ uniformly in $z\in [-\pi,\pi]$, where
$\phi_0$ satisfies
\begin{equation}
\label{limz} \phi_{0}'(z)=\frac{g\overline{J}}{2}w(\phi_0(z))
\end{equation}
and $\phi_0(0)=\pi$. Since, by (\ref{wz}), the constant function
$\pi$ is a solution to this initial-value problem, the uniqueness
theorem for initial-value problems implies that
$\phi_0(z)\equiv\pi$.

\vspace{0.4cm}
The following bounds, which follow immediately from (\ref{defrl})
and (\ref{mv}), will be useful:
\begin{lemma}
\label{ulb}

\noindent (i) We have, for all $z\in[-\pi,\pi]$:
$$\frac{1}{2\lambda}\rho_c(\lambda)\min_{x\in\Real}{J(x)}\leq R_{\lambda}(z)\leq
\frac{1}{2\lambda}\rho_c(\lambda)\max_{x\in\Real}{J(x)}.$$

\noindent (ii) If $J$ is not a constant function, the inequalities
of (i) are strict.
\end{lemma}

\begin{lemma}
\label{ell} In the excitable case $\beta<0$, we have
$$\frac{\lambda}{\rho_c(\lambda)}\geq \frac{g}{2|\beta|}\max_{x\in\Real}{J(x)}\;\;\Rightarrow\;\;
\Psi(\lambda)<1,$$
\end{lemma}

\noindent {\sc{proof:}} In the case of constant $J$ the result can
be proven by direct computation, so we now assume $J$ is not a
constant function. We will show that when
\begin{equation}\label{cl}
\frac{\lambda}{\rho_c(\lambda)}\geq \frac{g}{2|\beta|}\max_{x\in\Real}{J(x)}
\end{equation}
we have
\begin{equation}\label{cons1}
\phi_{\lambda}(z)<2\pi\;\;\;\forall z\in [0,\pi],
\end{equation}
\begin{equation}\label{cons2}
\phi_{\lambda}(z)>0 \;\;\;\forall z\in [-\pi,0].
\end{equation}
Together with (\ref{aps}), these imply the result of our lemma. To
prove our claim we note that, using
(\ref{tww}),(\ref{hz}),(\ref{wnn}), part (ii) of lemma \ref{ulb}
(which is why we need the assumption that $J$ is non-constant) and
(\ref{cl})
\begin{eqnarray}\label{kee}
\phi_{\lambda}(z)=0\;\;or\;\;2\pi \;&\Rightarrow\;&
\phi_{\lambda}'(z)=2\lambda\beta+2\lambda g
R_{\lambda}(z)\nonumber\\&<& 2\lambda\beta+g\rho_c(\lambda)\max_{x\in\Real}{J(x)}
\leq 2\lambda\beta +2\lambda|\beta|=0.
\end{eqnarray}
We now show that (\ref{kee}) implies (\ref{cons1}). If
(\ref{cons1}) fails to hold, then we set
$$z_0=\min \;\{z\in[0,\pi] \;|\; \phi_{\lambda}(z)= 2\pi\}.$$
This number is well-defined by continuity and by the fact that
$\phi_{\lambda}(0)=\pi$, which implies also that $z_0>0$. By
(\ref{kee}) we have $\phi_{\lambda}'(z_0)<0$, but this implies
that $\phi_{\lambda}(z)$ is decreasing in a neighborhood of $z_0$,
and in particular that there exist $z\in (0,z_0)$ satisfying
$\phi_{\lambda}(z)=2\pi.$ But this contradicts the definition
of $z_0$, and this contradiction proves (\ref{cons1}). Similarly,
assuming (\ref{cons2}) does not hold and defining
$$z_1=\max \;\{z\in [-\pi,0] \;|\; \phi_{\lambda}(z)=0\},$$
we conclude that $z_1<0$ and $\phi_{\lambda}'(z_1 )<0$, so that
$\phi_{\lambda}(z)$ is decreasing in a neighborhood of $z_1$, and
this implies a contradiction to the definition of $z_1$ and proves
that (\ref{cons2}) holds. This concludes the proof of the lemma.

\vspace{0.4cm}
Since
$$\lim_{\lambda\rightarrow\infty}{\rho_c(\lambda)}=e^{-\frac{c}{2}},$$
and $\rho_c(\lambda)$ is a monotone function so that
$$0<\rho_c(\lambda)\leq e^{-\frac{c}{2}}\;\;\;\forall \lambda>0,$$
we conclude from lemma \ref{ell} that

\begin{lemma}
\label{ell0} In the excitable case $\beta<0$, we have
$$\lambda\geq \frac{g e^{-\frac{c}{2}}}{2|\beta|}\max_{x\in\Real}{J(x)}\;\;\Rightarrow\;\;
\Psi(\lambda)<1,$$
\end{lemma}

Let us note that since $\Psi(\lambda)<1$ implies that (\ref{tt})
doesn't hold, and since $v=\frac{1}{\lambda}$, we can reformulate
the previous lemma as a lower bound for the velocities of rotating
waves in the excitable case.

\begin{lemma}\label{lbv}
In the excitable case $\beta<0$, we have the following lower bound
on the velocity of any rotating wave:
$$v > \frac{2|\beta|e^{\frac{c}{2}}}{\max_{x\in\Real}{J(x)}}\frac{1}{g}.$$
\end{lemma}

The following theorem shows that, in the excitable case and for
sufficiently weak synaptic coupling, there are no rotating waves
(so it implies part (II)(i) of theorem \ref{sum}).

\begin{theorem}
\label{noex} In the excitable case $\beta<0$, if $g\in (0,g_0)$,
where
$$g_0=\frac{\Omega(c,\beta)}{\max_{x\in\Real}{J(x)}},$$
with $\Omega(c,\beta)$ defined by (\ref{defom}),
then there exists no rotating wave.
\end{theorem}

\noindent {\sc{proof:}} We shall show that if $0<g<g_0$ then
\begin{equation}\label{sm}
\Psi(\lambda)<1
\end{equation}
for all $\lambda>0$, and thus that equation (\ref{tt}) cannot
hold. By (\ref{aps}), (\ref{sm}) is equivalent to
\begin{equation}
\label{din0} \phi_{\lambda}(\pi)-\phi_{\lambda}(-\pi)<2\pi.
\end{equation}
We note that, by lemma \ref{ell}, we already have (\ref{sm}) when
(\ref{cl}) holds, hence we may assume
\begin{equation}
\label{aaa} \lambda< \frac{g}{2|\beta|}\rho_c(\lambda)\max_{x\in\Real}{J(x)}.
\end{equation}
We define
$$\mu=\beta+\frac{g}{2\lambda}\rho_c(\lambda)\max_{x\in\Real}{J(x)},$$
and we note that (\ref{aaa}) is equivalent to the statement that
\begin{equation}
\label{mgz} \mu>0.
\end{equation}
Using (\ref{tww}) and lemma \ref{ulb} we have
\begin{eqnarray}\label{mmm}
\phi_{\lambda}'(z)&=&\lambda [h(\phi_{\lambda}(z))+ g
R_{\lambda}(z) w(\phi_{\lambda}(z))] \nonumber\\&\leq& \lambda
\Big[1-\cos(\phi_{\lambda}(x))+
\Big(\beta+\frac{g}{2\lambda}\rho_c(\lambda)\max_{x\in\Real}{J(x)}\Big)
(1+\cos(\phi_{\lambda}(z)))\Big]\nonumber\\&=& \lambda [(\mu
+1)+(\mu-1)\cos(\phi_{\lambda}(z))].
\end{eqnarray}
which implies (note that the integral below is well-defined
because of (\ref{mgz}))
$$\int_{-\pi}^{\pi}{\frac{\phi_{\lambda}'(z)dz}{(\mu
+1)+(\mu-1)\cos(\phi_{\lambda}(z))}}\leq 2\pi\lambda .$$ Making
the change of variables $\varphi=\phi_{\lambda}(z)$ we obtain
\begin{equation}\label{ha0}\int_{\phi_{\lambda}(-\pi)}^{\phi_{\lambda}(\pi)}
{\frac{d\varphi}{(\mu +1)+(\mu-1)\cos(\varphi)}}\leq 2\pi\lambda.
\end{equation}
If we assume, by way of contradiction, that (\ref{din0}) does not
hold, {\it{i.e.}} that
$\phi_{\lambda}(\pi)-\phi_{\lambda}(-\pi)\geq 2\pi$, then, using
(\ref{form}),
$$\int_{\phi_{\lambda}(-\pi)}^{\phi_{\lambda}(\pi)}{\frac{d\varphi}{(\mu
+1)+(\mu-1)\cos(\varphi)}}\geq
\int_{0}^{2\pi}{\frac{d\varphi}{(\mu
+1)+(\mu-1)\cos(\varphi)}}=\frac{\pi}{\sqrt{\mu}},$$ so together
with (\ref{ha0}) we obtain
$$\frac{1}{\sqrt{\mu}}\leq 2\lambda,$$
which is equivalent to
$$g\geq \frac{1}{\max_{x\in\Real}{J(x)}}\frac{1-4\beta\lambda^2}{2\lambda\rho_c(\lambda)},$$
which contradicts $g<g_0$. This contradiction proves
(\ref{din0}), concluding the proof of the theorem.

\vspace{0.4cm} We now proceed to prove that in the excitable case
when the synaptic coupling is sufficiently large we have at least
two rotating waves (see theorem \ref{ts} below).

\begin{lemma}
\label{two} In the excitable case $\beta<0$, if there exists some
$\lambda_0>0$ with
$$\Psi(\lambda_0)>1,$$
then there exist at least two solutions $\lambda_1,\lambda_2$ of
(\ref{tt}) with $0<\lambda_2<\lambda_0<\lambda_1$, hence
two rotating waves, with velocities satisfying
$$v_1=\frac{1}{\lambda_1}<\frac{1}{\lambda_0}<\frac{1}{\lambda_2}=v_2.$$
\end{lemma}

\noindent {\sc{proof:}} By lemma \ref{psiprop}, we can choose
$\lambda'_2<\lambda_0$ so that $\Psi(\lambda'_2)<1$. By
lemma \ref{ell0}, if we fix
$$\lambda'_1=\frac{g e^{-\frac{c}{2}}}{2|\beta|}\max_{x\in\Real}{J(x)},$$
then $\Psi(\lambda)<\frac{1}{\lambda}$ for all $\lambda\geq
\lambda'_1$, and in particular it follows that
$\lambda'_1>\lambda_0$. We thus have
$$\lambda'_2<\lambda_0<\lambda'_1$$
with
$$\Psi(\lambda'_2)<1,\;\;\Psi(\lambda_0)>1,\;\;
\Psi(\lambda'_1)<1.$$ Thus by the intermediate value theorem, the
equation (\ref{tt}) has a solution $\lambda_2\in(\lambda'_2,\lambda_0)$
and a solution $\lambda_1\in(\lambda_0,\lambda'_1)$,
corresponding to two rotating waves.

\vspace{0.4cm}
The following lemma is valid for all values of $\beta$:

\begin{lemma}\label{glar}
Assume that $\lambda>0$ satisfies the inequality
\begin{equation}\label{kein}
f_{c,\beta}(\lambda)<g\min_{x\in\Real}{J(x)},
\end{equation}
where $f_{c,\beta}$ is defined by (\ref{df}). Then
$$\Psi(\lambda)>1.$$
\end{lemma}

\noindent {\sc{proof:}} By (\ref{aps}), our claim is equivalent to
\begin{equation}
\label{din}
\phi_{\lambda}(\pi)-\phi_{\lambda}(-\pi)>2\pi.
\end{equation}
We define
$$\eta=\beta+\frac{g}{2\lambda}\rho_c(\lambda)\min_{x\in\Real}{J(x)},$$
and we note that (\ref{kein})
is equivalent to
\begin{equation}
\label{eg2}\eta>\frac{1}{4\lambda^2}.
\end{equation}
Using (\ref{tww}) and lemma \ref{ulb} we have
\begin{eqnarray}\label{lll}
\phi_{\lambda}'(z)&=&\lambda [h(\phi_{\lambda}(z))+ g
R_{\lambda}(z) w(\phi_{\lambda}(z))] \nonumber\\&\geq&
\lambda\Big[1-\cos(\phi_{\lambda}(x))+
\Big(\beta+\frac{g}{2\lambda}\rho_c(\lambda)\min_{x\in\Real}{J(x)}\Big)
(1+\cos(\phi_{\lambda}(z)))\Big]\nonumber\\&=& \lambda [(\eta
+1)+(\eta-1)\cos(\phi_{\lambda}(z))],
\end{eqnarray}
which implies
$$\int_{-\pi}^{\pi}{\frac{\phi_{\lambda}'(z)dz}{(\eta
+1)+(\eta-1)\cos(\phi_{\lambda}(z))}}\geq 2\pi\lambda .$$ Making
the change of variables $\varphi=\phi_{\lambda}(z)$, we obtain
\begin{equation}\label{ha}\int_{\phi_{\lambda}(-\pi)}^{\phi_{\lambda}(\pi)}
{\frac{d\varphi}{(\eta
+1)+(\eta-1)\cos(\varphi)}}\geq 2\pi\lambda.
\end{equation}
If we assume, by way of contradiction, that (\ref{din}) does not
hold, {\it{i.e.}} that $\phi_{\lambda}(\pi)-\phi_{\lambda}(-\pi)\leq 2\pi$,
then, using (\ref{form}),
$$\int_{\phi_{\lambda}(-\pi)}^{\phi_{\lambda}(\pi)}{\frac{d\varphi}{(\eta
+1)+(\eta-1)\cos(\varphi)}}\leq
\int_{0}^{2\pi}{\frac{d\varphi}{(\eta
+1)+(\eta-1)\cos(\varphi)}}=\frac{\pi}{\sqrt{\eta}},$$ so together
with (\ref{ha}) we obtain
$$\frac{1}{\sqrt{\eta}}\geq 2\lambda.$$
This contradicts (\ref{eg2}), and this contradiction implies that
(\ref{din}) holds, completing our proof.
\vspace{0.4cm}

The following theorem implies part (II)(ii) of theorem \ref{sum}.

\begin{theorem}
\label{ts} In the excitable case $\beta<0$, let
\begin{equation}\label{defg2}
g_1=\frac{\Omega(c,\beta)}{\min_{x\in\Real}{J(x)}},
\end{equation}
where $\Omega(c,\beta)$ is defined by (\ref{defom}).
Then when $g>g_1$, there exist at least two rotating waves. In
fact, we have a `slow' wave with velocity $v_s$ bounded from above
by
\begin{equation}\label{sw}
v_s\leq \underline{v}_{c,\beta}\Big(g\min_{x\in\Real}{J(x)}\Big)
\end{equation}
and a `fast wave' with velocity $v_f$ bounded from below by
\begin{equation}\label{fw}
v_f\geq \overline{v}_{c,\beta}\Big(g\min_{x\in\Real}{J(x)}\Big),
\end{equation}
where $\underline{v}_{c,\beta},\overline{v}_{c,\beta}$ are the functions defined
by (\ref{vdef}).

\noindent
As a consequence of (\ref{sw}),(\ref{fw}) we have, for the slow
wave
\begin{equation}
\label{swe} v_s\leq
\frac{2|\beta|e^{\frac{c}{2}}}{\min_{x\in\Real}{J(x)}}\frac{1}{g}+O\Big(\frac{1}{g^3}\Big)
\;\;\;as\;\;g\rightarrow\infty,
\end{equation}
for the fast wave in the case $c>0$
\begin{equation}\label{fwe0}
v_f\geq 2\sqrt{\frac{\pi\min_{x\in\Real}{J(x)}}{e^{\frac{c}{2}}-1}}\sqrt{g}+O(1)
\;\;\;as\;\;g\rightarrow\infty,
\end{equation}
and for the fast wave in the case $c=0$
\begin{equation}
\label{fwe}
v_f\geq 2\min_{x\in\Real}{J(x)}g+O\Big(\frac{1}{g}\Big)
\;\;\;as\;\;g\rightarrow\infty.
\end{equation}
\end{theorem}

\noindent {\sc{proof:}}
$g>g_1$ and (\ref{defom}) imply the existence of $\lambda>0$
satisfying (\ref{kein}), hence by lemma \ref{glar} $\Psi(\lambda)>1$,
so that lemma \ref{two} implies the existence of two rotating
waves.

To prove (\ref{sw}),(\ref{fw}), we note that, assuming $g>g_1$,
the range of values of $\lambda_0$ for which (\ref{kein}) holds is
the interval
$$\underline{\lambda}_{c,\beta}(g\min_{x\in\Real}{J(x)})<\lambda_0<
\overline{\lambda}_{c,\beta}(g\min_{x\in\Real}{J(x)}),$$ where the
functions
$\underline{\lambda}_{c,\beta},\overline{\lambda}_{c,\beta}$ are
defined in section \ref{constant}. Thus, applying lemma \ref{two}
with
$$\lambda_0=\overline{\lambda}_{c,\beta}(g\min_{x\in\Real}{J(x)})-\epsilon,$$
where $\epsilon>0$ is arbitrarily small, we obtain the existence
of a solution $\lambda_{\epsilon}$ of (\ref{tt}) with
$\lambda_{\epsilon}>\overline{\lambda}_{c,\beta}(g\min_{x\in\Real}{J(x)})-\epsilon$.
Since $\epsilon>0$ is arbitrary, we have a solution $\lambda$ of
(\ref{tt}) with $\lambda\geq
\overline{\lambda}_{c,\beta}(g\min_{x\in\Real}{J(x)})$, hence a
rotating wave with velocity $v_s$ satisfying (\ref{sw}). Similarly
applying lemma \ref{two} with
$\lambda_0=\underline{\lambda}_{c,\beta}(g\min_{x\in\Real}{J(x)})+\epsilon$,
we obtain the existence of a wave with velocity $v_f$ satisfying
(\ref{fw}).

The estimates (\ref{swe})-(\ref{fwe}) follow from
(\ref{sw}),(\ref{fw}) and lemma \ref{asa}.

\vspace{0.4cm} We note that along with the upper bound
(\ref{swe}), we have a lower bound for the velocity of the slow
wave, given by lemma \ref{lbv}.

\begin{question}
Derive an upper bound for the velocities of the fast waves (note that
(\ref{fw}) gives a lower bound).
\end{question}

\begin{question}
Theorems \ref{noex} and \ref{ts} show that several of the
qualitative features that we saw explicitly in the case of uniform
coupling (section \ref{constant}) remain valid in the general
case. It is natural to ask whether more can be said, {\it{e.g.}},
whether the following conjecture, or some weakened form of it, is
true: for any $J$, there exists a value $g_{crit}$ such that:

\noindent (i) For $g<g_{crit}$ there exist no travelling waves.

\noindent (ii) For $g=g_{crit}$ there exists a unique travelling wave.

\noindent (iii) For $g>g_{crit}$ there exist {\it{precisely}} two travelling
waves.
\end{question}

The next theorem deals with the oscillatory case $\beta>0$, as
well as the borderline case $\beta=0$, and in particular proves
part (I) of theorem \ref{sum}.

\begin{theorem}\label{osct}
\label{osc} If $\beta\geq 0$, there exists a rotating wave
solution for any value of $g>0$, with velocity $v$ bounded from
below by
\begin{equation}\label{fw1}
v\geq v(g\min_{x\in\Real}{J(x)}),
\end{equation}
where $v$ is the function defined by (\ref{vdef1}), and
the asymptotic formulas (\ref{fwe0}),(\ref{fwe}) hold with $v_f$
replaced by $v$.
\end{theorem}

\noindent {\sc{proof:}} If $\beta>0$, then for any $g>0$ the equation
$$f_{c,\beta}(\lambda)=g\min_{x\in\Real}{J(x)}$$
has the unique solution
$\lambda_{c,\beta}(g\min_{x\in\Real}{J(x)}).$  Hence, any
$$\lambda_0>\lambda_{c,\beta}(g\min_{x\in\Real}{J(x)})$$
satisfies (\ref{kein}), so that by lemma \ref{glar}
$\Psi(\lambda_0)>1$. On the other hand for $\lambda$ sufficiently
small we have, by lemma \ref{psiprop}, $\Psi(\lambda)<1$. Hence
there exists a solution $\lambda\in (0,\lambda_0)$ of (\ref{tt}).
Since $\lambda_0>\lambda_{c,\beta}(g\min_{x\in\Real}{J(x)})$ is
arbitrary, we conclude that there exists a solution $\lambda\leq
\lambda_{c,\beta}(g\min_{x\in\Real}{J(x)})$ of (\ref{tt}). Hence a
rotating wave with velocity satisfying (\ref{fw1}).

\begin{question}
Is it true that in the oscillatory case $\beta\geq 0$ the rotating
wave is always unique? We saw that this is the case when $J\equiv
1$.
\end{question}

Finally, we stress the important question of stability of the
rotating waves, which remains open:

\begin{question}
Investigate the question of stability of the rotating waves,
{\it{i.e.}}, do arbitrary solutions of (\ref{ge}), (\ref{syn})
approach one of the rotating waves in large time? We conjecture
that, at least under some restrictions on $J$, the rotating wave
is stable in the case $\beta>0$, while in the case $\beta<0$ the
fast rotating wave is stable and the slow one is unstable.
\end{question}

\end{document}